\documentclass{article}
\usepackage{arxiv}
\usepackage[utf8]{inputenc} % allow utf-8 input
\usepackage[T1]{fontenc}    % use 8-bit T1 fonts
\usepackage{url}            % simple URL typesetting
\usepackage{booktabs}       % professional-quality tables
\usepackage{amsfonts}       % blackboard math symbols
\usepackage{nicefrac}       % compact symbols for 1/2, etc.
\usepackage{microtype}      % microtypography
\usepackage{graphicx}
\usepackage{natbib}
\usepackage{doi}
\usepackage{array}
\usepackage{multirow}
\usepackage{mathtools}
\usepackage{amsmath}
\usepackage{adjustbox}
\usepackage{afterpage}
\usepackage{icomma}
\usepackage{nicematrix}
\usepackage{xltabular}
\usepackage{caption}
\usepackage{subcaption}
\usepackage{float}

\usepackage{xr-hyper}

\makeatletter
\newcommand*{\addFileDependency}[1]{%
  \typeout{(#1)}%
  \@addtofilelist{#1}%
  \IfFileExists{#1}{}{\typeout{No file #1.}}%
}
\makeatother

\makeatletter
\renewcommand\normalsize{%
  \@setfontsize\normalsize{11}{14.5} 
  \abovedisplayskip      7\p@ \@plus 2\p@ \@minus 5\p@
  \abovedisplayshortskip \z@ \@plus 3\p@
  \belowdisplayskip      \abovedisplayskip
  \belowdisplayshortskip 4\p@ \@plus 3\p@ \@minus 3\p@
}
\normalsize
\makeatother
\usepackage{setspace}

\newcommand{\f}{_\text{CE}}
\newcommand{\re}{_\text{RE}}
\newcommand{\m}{_\text{ME}}
\newcommand{\aic}{\text{AIC}}
\newcommand{\qt}{Q_{\text{total}}}
\newcommand{\qh}{Q_{\text{het}}}
\newcommand{\qi}{Q_{\text{inc}}}

\title{Identifying Conditions Favouring Multiplicative Heterogeneity Models in Network Meta-Analysis}

\author{ 
\href{https://orcid.org/0009-0000-7290-1316}{\includegraphics[scale=0.06]{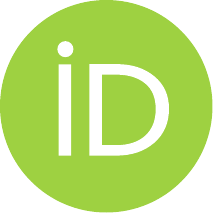} \hspace{1mm}Xinlei Xu} \\
\texttt{xinlei.xu@uwaterloo.ca}
\and \href{https://orcid.org/0000-0002-8272-797X}{\includegraphics[scale=0.06]{orcid.pdf}\hspace{1mm}Caitlin H. Daly}\\      
\texttt{caitlin.daly@uwaterloo.ca} 
\and \href{https://orcid.org/0000-0003-4124-2498}{\includegraphics[scale=0.06]{orcid.pdf}\hspace{1mm}Audrey B{\'e}liveau}\\
\texttt{audrey.beliveau@uwaterloo.ca}\\
[3pt] Department of Statistics and Actuarial Science, University of Waterloo \\Waterloo, Ontario, N2L 3G1, Canada }

\date{}

\renewcommand{\headeright}{}
\renewcommand{\undertitle}{}
\renewcommand{\shorttitle}{Conditions Favouring Multiplicative Heterogeneity in NMA}

\hypersetup{
pdftitle={Identifying Conditions Favouring Multiplicative Heterogeneity Models in Network Meta-Analysis},
pdfauthor={Xinlei Xu, Caitlin H Daly, Audrey B{\'e}liveau},
pdfkeywords={network meta-analysis; heterogeneity; random‐effects; weighted least squares},
}

\begin{document}
\maketitle

\vspace{-1cm}
%TC:ignore
\section*{Acknowledgements}

This study uses data obtained from publicly accessible databases. All datasets used in the analyses are available from the \texttt{nmadb} R package. This research was supported by A. Béliveau's NSERC Discovery Grant (RGPIN-2019-04404).
%TC:endignore
\vspace{11pt}

\begin{spacing}{1.5}
\begin{abstract}

Explicit modelling of between‐study heterogeneity is essential in network meta‐analysis (NMA) to ensure valid inference and avoid overstating precision. While the additive random‐effects (RE) model is the conventional approach, the multiplicative‐effect (ME) model remains underexplored. The ME model inflates within‐study variances by a common factor estimated via generalized least squares, yielding identical point estimates to a common‐effect model while inflating confidence intervals. We empirically compared RE and ME models across NMAs of two-arm studies with significant heterogeneity from the \texttt{nmadb} database, assessing model fit using the Akaike Information Criterion. The ME model often provided comparable or better fit to the RE model. Case studies further revealed that RE models are sensitive to extreme and imprecise observations, whereas ME models assign less weight to such observations and hence exhibit greater robustness to publication bias. Our results suggest that the ME model warrant consideration alongside conventional RE model in NMA practice.
\end{abstract}

\keywords{network meta-analysis\and heterogeneity\and random‐effects\and weighted least squares}

\end{spacing}

\begin{spacing}{2}
\section{Introduction} \label{s:intro}

Meta-analysis (MA) is a statistical technique for combining results from studies that compare two treatments to address the same research question, yielding an overall estimate of effect size \citep{borenstein2021introduction}. It is widely applied across various scientific disciplines—such as psychology and ecology—while predominantly used in the health field to synthesize findings from randomized clinical trials (RCT). Statistically combining relative effects from RCTs preserves randomization within trials, thereby facilitating an unbiased and more precise comparison of treatment effects, assuming there are no systematic differences between trials \citep{ades2024twenty}.

Network meta-analysis (NMA) extends this framework to compare multiple treatments simultaneously \citep{lumley2002network}. By assuming systematic differences in effect sizes between studies arise exclusively from different treatments being implemented, NMA can capitalize on the consistency of relative effects, $d_{a,b}=d_{c,b}-d_{c,a}$, for any three interventions $a, b$ and $c$, with $d_{a,b}$ denoting the relative treatment effect of $b$ relative to $a$ \citep{schwarzer2015meta}. NMA integrates both direct and indirect evidence to estimate all pairwise comparisons including those never directly investigated in any single study, and enables treatment ranking \citep{dias2019network}. 

Between-study heterogeneity arises when there are differences in effect modifiers across studies, leading to some variation in the relative effects estimated by each study. While health researchers strive to identify a set of homogeneous studies in terms of their population, interventions (e.g., dose, route of administration, administrator), effect measure definitions (e.g., measurement tool, timing of measurement), and other settings (e.g., acute vs community care), some between-study heterogeneity may be tolerated and modelled in order to address research gaps; unintended between-study heterogeneity may also be present due to limited knowledge of effect modifiers. Explicitly modelling between-study heterogeneity in both MA and NMA avoids overstating the precision of pooled estimates and ensures valid inferences for the target population when true effects vary across settings. Appropriately accounting for heterogeneity reflects the reality that studies differ in populations, interventions, outcomes and designs. Ignoring heterogeneity can yield overconfident inferences and misleading results \citep{thompson1999explaining}. The conventional approach to account for heterogeneity in MA and NMA is the random-effects (RE) model, which introduces an additive between-study variance component, $\tau^2$, assuming constant dispersion across treatment comparisons.

In the context of pairwise MA, \citet{thompson1999explaining} considered an alternative approach to modelling heterogeneity that employs a multiplicative parameter that inflates the within-study variances. This multiplicative parameter can be estimated through unrestricted weighted least squares (UWLS) rather than conventional RE likelihood-based methods. Building on this idea, \citet{Kulinskaya2014_overdispersion} proposed a general overdispersion framework that unifies additive and multiplicative formulations by introducing a variance inflation factor as an intra-class correlation parameter, thereby embedding the traditional RE model as a special case and allowing both over- and under-dispersion in the data. Subsequent research found that smaller studies tend to be systematically more heterogeneous, violating the assumptions of the RE model, and that the UWLS estimator provides a more accurate and robust alternative for meta-analysis \citep{stanley2015neither, stanley2022beyond}. Large-scale empirical evaluations further demonstrated that UWLS often fits observed data similar to or better than RE models, as indicated by smaller AIC or BIC values \citep{mawdsley2017accounting,stanley2023unrestricted}. The UWLS framework was later extended to accommodate meta-regression \citep{stanley2017neithermr}. Finally, \citet{schmid2017heterogeneity} proposed a hybrid variance structure combining both additive and multiplicative components of heterogeneity, offering a more flexible way to model heterogeneity.

While multiplicative effects models have been extensively studied in the context of pairwise MA; they have received limited attention in the context of NMA. \citet{tu2017using} incorporated the UWLS approach in a contrast-based structural equation model (SEM) as a multiplicative variance inflation per pairwise comparison. Building on this idea, \citet{shih2019evaluating} embedded the UWLS framework in the arm-parameterized SEM and demonstrated improved model fit in two published NMAs. More recently, \citet{evrenoglou2022network} modelled the heterogeneity multiplicatively within a penalized-likelihood regression model for NMA, showing improved bias reduction and proper coverage relative to conventional methods used for NMA of rare events through a simulation study. 

In much of the existing literature developing multiplicative effects models, their use was motivated by the analytical and computational challenges associated with generalized linear mixed models, whose likelihood functions for the RE model are typically intractable and require approximation methods. In contrast, an important distinction of our work is its investigation of multiplicative effect (ME) models within the classical normal-normal NMA framework, where the likelihood under the standard RE model has a closed-form expression. Our motivation for introducing a multiplicative variance structure in this setting is thus not driven by likelihood intractability, but rather by the potential for improved modelling of heterogeneity.

There is currently a lack of empirical evidence comparing the fit of ME and RE models in the NMA setting, and discussion on the conditions affecting the comparative performance of these models. In this paper, we address this gap by conducting a large-scale empirical comparison of the ME model versus the RE model using an open-access database of NMAs including at least four different interventions in RCTs published up to April 14, 2015 \citep{petropoulou2017bibliographic, Papakonstantinou2019nmadb}. The remainder of the paper is organized as follows. Section~\ref{s:methods} outlines the methodology, beginning with model definitions, followed by model comparison procedures and our empirical approach. Section~\ref{s:results} presents our findings including case study analyses of specific NMA datasets that exhibit significant discrepancies in model performance. Finally, Section~\ref{s:discussion} discusses practical implications, limitations, and directions for future research.

\section{Preliminaries} \label{s: preliminaries}

We focus on frequentist contrast-likelihood models of NMA under the assumption of consistency \citep{rucker2012network,harrer2021doing}. We start by presenting models for NMAs of two-arm studies only, as these are conceptually simpler. Let $\mathbf{y} = (y_1,....,y_m)^\top$ be the vector of estimated relative treatment effects in studies $1,...,m$ and $\mathbf V = \text{diag}(s_1^2,...,s_m^2)$ be the corresponding diagonal variance matrix, where $s_1,...,s_m$ are assumed to be known although in practice they are the estimated standard
errors reported within each study. We assume the data form a connected network of $n$ treatments. To parametrize the consistency assumption, let $\mathbf d \in \mathbb R^{n-1}$ denote a vector of underlying mean relative effects, each defined with respect to a common reference treatment (typically placebo or control), such that the relative effect between any two non‑reference treatments can be obtained as the difference between their corresponding basic parameters. Let $\mathbf X\in \mathbb R^{m\times (n-1)}$ denote the design matrix utilizing -1, 1, and 0 to encode the treatment contrasts implemented in each study, such that $\mathbb{E}(\mathbf y)=\mathbf X\mathbf d$. 

Under the common-effect (CE) NMA model, $\mathbf{y}=\mathbf X\mathbf d+\boldsymbol{\varepsilon}$, with $ \boldsymbol\varepsilon\sim MVN(\mathbf0, \mathbf V)$. The weighted least squares (WLS) estimator of $\mathbf d$ is $\mathbf{\hat d}\f=(\mathbf X^\top\mathbf V^{-1}\mathbf X)^{-1}\mathbf X^\top\mathbf V^{-1}\mathbf y$ with $\text{Var}(\mathbf{\hat d}\f)=(\mathbf X^\top\mathbf V^{-1}\mathbf X)^{-1}$. To account for between-study heterogeneity, the conventional random-effects (RE) model augments the within-study variances by an additive variance component, $\tau^2$: 
$$\mathbf{y}=\mathbf X\mathbf d+\boldsymbol{\varepsilon}, \quad \boldsymbol\varepsilon\sim MVN(\mathbf 0, \mathbf \Sigma =\mathbf V+\tau^2 \mathbf I),$$
where $\mathbf I$ is the $m\times m$ identity matrix. The corresponding WLS estimator is $\mathbf{\hat d}\re=(\mathbf X^\top\boldsymbol{\hat\Sigma}^{-1}\mathbf X)^{-1}\mathbf X^\top\boldsymbol{\hat\Sigma}^{-1}\mathbf y$ with $\boldsymbol{\hat\Sigma}=\mathbf V+\hat\tau^2 \mathbf I$ and the variance of $\mathbf{\hat d}\re$ can be estimated by $(\mathbf X^\top\boldsymbol{\hat\Sigma}^{-1}\mathbf X)^{-1}$. The additive heterogeneity parameter estimate $\hat\tau^2$ is commonly obtained by restricted maximum likelihood (REML) \citep{white2015nma} or the DerSimonian-Laird (DL) method given by $\hat\tau^2_\text{DL}=\max\left\{0,\frac{\qt - df_\text{total}}{\sum_{i=1}^{m}(1-h_{i})/s_i^2}\right\}.$ Here, 
$\qt = (\mathbf y-\mathbf{\hat y}^\text{CE})^\top\mathbf V^{-1} (\mathbf y-\mathbf{\hat y}^\text{CE})$ is the total $Q$ statistic computed under the CE model, with fitted values $\mathbf{\hat y}^\text{CE} = \mathbf X \mathbf{\hat d}\f$,
$df_\text{total} = m - (n-1)$, and $h_i$ is the leverage of study $i$ in the CE model, formally defined in Appendix \ref{appen_proof} \citep{rucker2014reduce, dersimonian1986meta}.

Alternatively, the multiplicative‐effect (ME) model for NMA \citep{shih2019evaluating, evrenoglou2022network} scales each within‐study variance by a common factor $\phi$:
$$\mathbf{y}=\mathbf X\mathbf d+\boldsymbol{\varepsilon}, \quad \boldsymbol\varepsilon\sim MVN(\mathbf0, \phi\mathbf{V}).$$

The WLS estimate is identical to that of the CE model, $\mathbf{\hat d}\m=(\mathbf X^\top\mathbf V^{-1}\mathbf X)^{-1}\mathbf X^\top\mathbf V^{-1}\mathbf y = \mathbf{\hat d}\f$, but its variance is inflated by $\phi$, $Var(\mathbf{\hat d}\m)=\phi(\mathbf X^\top\mathbf V^{-1}\mathbf X)^{-1}=\phi \text{Var}(\mathbf{\hat d}\f)$. As a result, confidence intervals (CIs) for elements of $\mathbf{\hat d}\m$ or for predictions of individual $y_i$ values are simply scaled versions of the CE model CIs (at the same confidence level), with their widths inflated by a factor of $\hat\phi$.

A key practical advantage of the ME model is that it can be fit and diagnosed using standard multiple linear regression techniques. This is because a linear transformation of the response, $\mathbf y^*=\mathbf V^{-1/2}\mathbf y$, yields a linear regression model with homoskedastic errors, $\mathbf y^* = \mathbf X^*\mathbf d + \boldsymbol\epsilon^*$, where $\mathbf X^* = \mathbf V^{-1/2}\mathbf X\mathbf d$ and $\boldsymbol\epsilon^*\sim MVN(0, \phi \mathbf I).$  Consequently, weighted least squares applied to the original model is equivalent to ordinary least squares applied to the transformed model. Moreover, $\phi$ admits a minimum-variance unbiased estimator given by the standard Pearson-type residual mean square \citep{seber2003linear}, obviating the need for REML or ad hoc plug-in approaches. A truncated version of the estimator is:
\begin{equation*}
    \hat{\phi} = \max\left\{1, \frac{1}{m - (n - 1)} (\mathbf y^* -\mathbf X^*\mathbf {\hat{d}}\f)^\top \mathbf  (\mathbf y^* - \mathbf X^*\mathbf {\hat{d}}\f)\right\} \\
    = \max\left\{1, \frac{\qt}{df_\text{total}} \right\}.
\end{equation*}
Here, the degrees of freedom at the denominator are obtained in the usual way by subtracting the number of estimated basic parameters from the number of observations. The constraint $\hat\phi\geq 1$ avoids deflating study variances to ensure that heterogeneity is not underestimated. Note that $\hat\phi$ is also equivalent to taking the reciprocal of the complement of the generalized $I^2$ statistic: $\hat\phi = \frac{1}{1-I^2}$, which makes its interpretation as a variance inflation factor transparent \citep{rucker2014reduce}. Finally, because both $\hat\phi$ and $\hat \tau^2_\text{DL}$ are rescaling of $Q_\text{total}$, we can show that $\hat\phi = 1$ if and only if $\hat\tau^2_\text{DL} = 0$ (see Appendix~\ref{appen_proof}); this equivalence does not generally hold under REML estimation. %can be shown that $\hat\phi = 1$ if and only if $\hat\tau = 0$ (see Appendix~\ref{appen_proof}).

In NMAs that include multi‑arm studies, the vector $\mathbf{y}$ and matrix $\mathbf V$ increase in dimension, with $\mathbf V$ taking a block‑diagonal form to represent within-study covariances. Estimation of the ME model then proceeds via generalized least squares, rather than WLS, to accommodate the off-diagonal elements in $\mathbf V$. Importantly, the estimation formulas remain unchanged apart from using the augmented $\mathbf{y}$ and $\mathbf V$, while the core principle of multiplying $\mathbf V$ (including covariances) by $\phi$ continues to apply. This follows because multiplying within‑study variances by $\phi$ also multiplies the variance of their differences by the same factor, implying that the corresponding covariances in $\mathbf V$ must likewise be scaled by $\phi$. 

We further note that the ME model appears to extend naturally to a range of more complex NMA settings. While a full investigation is beyond the scope of this paper, we do not foresee technical obstacles to replacing additive variance structures with multiplicative ones in a variety of existing frameworks -- for example, inconsistency models, additive component NMA models, and meta‑regression models. Decomposition of the pooled estimates into direct and indirect components for assessing local inconsistency would follow the same procedure as in CE and RE models, with the only difference being that the variances of the direct and indirect CE estimates are inflated by $\hat\phi$.

\section{Methods} \label{s:methods}

To identify the conditions affecting the comparative performance of the ME and RE models, we carried out an empirical evaluation comparing model fit across a large number of real-world NMA examples. Our data source was \texttt{nmadb}, a large database of NMAs published between 1999 and 2015 \citep{petropoulou2017bibliographic, Papakonstantinou2019nmadb}. This database contained 453 NMAs, of which 262 were identified with available data. We restricted our attention to NMAs of two-arm studies, which are more common than multi-arm studies. Although multi-arm studies would be of methodological interest, characteristics of the \texttt{nmadb} data currently makes their investigation challenging; see Section \ref{appen_multi_arm} for further discussion.

We focused specifically on odds ratio (OR), risk ratio (RR), and mean differences (MD) because these three effect measures are the most common summary effect measures in NMA and in \texttt{nmadb} \citep{zarin2017characteristics}. Restricting attention to these effect measures maximized the number of networks available within each effect measure category, improving the reliability of comparisons. For each dataset within each effect measure type, both the RE model and the ME model were fitted.

To compare model fit, we computed Akaike's Information Criterion (AIC), defined as $\aic = 2k - 2\log \hat{L}$, where $\hat{L}$ is the maximized likelihood and $k$ is the number of estimated parameters \citep{akaike1998information}. Since the ME and RE models have the same complexity, this is equivalent to comparing the log-likelihoods. We computed $\Delta \aic = \aic\m-\aic\re$, so that negative $\Delta\aic$ suggests that the ME model is favoured.  As a rule of thumb, $|\Delta\aic| \leq 3$ suggests that both models are similarly supported by the data, and $|\Delta\aic| > 3$ indicates a clear preference for one model \citep{burnham2002model}. Since the RE and ME models have the same number of estimated parameters, $n-1$ relative effects and one heterogeneity parameter, the difference in AIC can be expressed as
\begin{align*}
 \Delta\aic= &\log \det (\hat\phi\mathbf  V) + (\hat{\mathbf y} - \mathbf{X}\hat{\mathbf d}_\text{ME})^\top (\hat\phi\mathbf  V)^{-1} (\hat{\mathbf y} - \mathbf{X}\hat{\mathbf d}_\text{ME}) \\&- \log \det(\mathbf V+\hat\tau^2 \mathbf I) - (\hat{\mathbf y} - \mathbf{X}\hat{\mathbf d}_\text{RE})^\top (\mathbf V+\hat\tau^2 \mathbf I)^{-1} (\hat{\mathbf y} - \mathbf{X}\hat{\mathbf d}_\text{RE}).
\end{align*}

Within each effect measure type, we investigated whether $|\Delta\aic|$ was reflective of the amount of total heterogeneity in the NMA. In each NMA, we calculated the test statistic $\qt$, which measures lack of fit for the CE model. Despite the historical use of this statistic when discriminating between CE and RE models, this statistic actually does not posit or imply an alternative, better-fitting model to the CE model, such as one with additive or multiplicative heterogeneity. We calculated the $p$-value using $\qt$ as the test statistic for testing the null hypothesis $H_0:$ there is no between-study heterogeneity or inconsistency, versus the alternative $H_A:$ there is between-study heterogeneity or inconsistency. Under $H_0$, the RE and ME models are equivalent as they both reduce to the CE model, and $\qt\sim \chi^2(df_\text{total})$. Below the 10 \% significance level \citep{cochrane2024chapter10}, small $p$-values raise the concern that the RE and ME models may not be similar in their fit to the data, producing larger $|\Delta\aic|$ values.

The NMAs with extreme $|\Delta\mathrm{AIC}|$ (>9) warrant closer investigation. Such large departures indicate that one heterogeneity model is strongly favoured by the data. We therefore investigate targeted case studies \citep{jansen2014comparative, mason2004topical, cooper2012network, brodszky2011rheumatoid, biondi2013comparative}. By examining NMAs with strong preference for either the RE or ME model, we aimed to uncover the structural and statistical characteristics (e.g., network structure, magnitude of between‐study dispersion, impact of small studies) that drive model selection. 

For each case study, we examined the topology of the treatment network, computed $\qt$ and further decomposed it into the heterogeneity part and inconsistency part \citep{krahn2013graphical}. Let $C$ be the number of unique direct pairwise comparisons (study designs) in an NMA and $S_c$ be the collection of studies contributing to comparison $c=1,...,C$. We decomposed $\qt = \qh+\qi$, where 
$$Q_{\text{het}}=\sum_{c=1}^C\sum_{i\in S_c}w_i(y_i-\bar y_c)^2, \quad \text{with }\bar y_c = \frac{\sum_{i\in S_c}w_iy_i}{\sum_{i\in S_c}w_i}, \quad w_i=\frac{1}{s_i^2},$$ 
measures the heterogeneity between studies in the same study designs in a CE model, and the inconsistency component
$$Q_{\text{inc}}=\sum_{c=1}^C\sum_{i\in S_c}w_i(\bar y_c - \hat y^\text{CE}_i)^2$$ 
captures the heterogeneity between direct and indirect evidence in the CE model \citep{lu2006assessing, higgins2012consistency}. Under $H_0$, each component follows a chi-squared distribution: $Q_{\text{het}}\sim \chi^2(m-C), Q_{\text{inc}}\sim \chi^2(C-(n-1)),$ and we respectively calculate the $p$-values for testing the null hypotheses of no between-study heterogeneity and no inconsistency. 

Moreover, to understand the causes of discrepancies between the RE and FE model, we calculated the contribution to $\qh$ from design $c$, $\qh^c=\sum_{i\in S_c}\qh^i$, where $\qh^i=w_i(y_i-\bar y_c)^2$ is the contribution from study $i$. These were displayed on forest plots where we also compared estimated pooled relative effects under the ME and RE models with the estimated relative effects in each study (${\mathbf y}$) along with their associated CIs.

We conducted the analysis using the R software (Version 4.5.1). The \texttt{nmadb} database was accessed via the \texttt{nmadb} R package \citep{Papakonstantinou2019nmadb, petropoulou2017bibliographic}. The CE and RE models were fitted via the R package \texttt{netmeta} and its eponymous function \citep{balduzzi2023netmeta}. In RE models, the additive heterogeneity parameter $\tau^2$ was estimated by the DL and REML methods to enable their comparison. The ME model was obtained by inflating the variance of the CE model estimates from \texttt{netmeta} by $\hat\phi$.

\section{Results} \label{s:results}

Among the 120 NMAs including only two-arm studies in \texttt{nmadb}, 48 reported odds ratios (OR), 38 used risk ratios (RR), and 14 employed mean differences (MD) (in total, 100 networks were considered for the analysis). In this section, we focus on results based on $\hat\tau^2_\text{DL}$. Corresponding results under the REML estimator for $\tau^2$ are presented in Appendix~\ref{subsec_suppresults} and are largely consistent with the main findings, with any discrepancies discussed therein.

%Overall, estimating $\tau$ by REML showed broadly similar qualitative patterns, although the distributions of $\Delta \aic$ (Figure~\ref{fig:AICdiff_REML}) differed somewhat from those obtained under DL estimation. In particular, the REML estimator yielded much fewer cases with $\Delta\aic = 0$ than the DL estimator, which is further discussed in Appendix~\ref{subsec_suppresults}.

\begin{figure}[htbp]
 \centering
 \includegraphics[width=\textwidth]{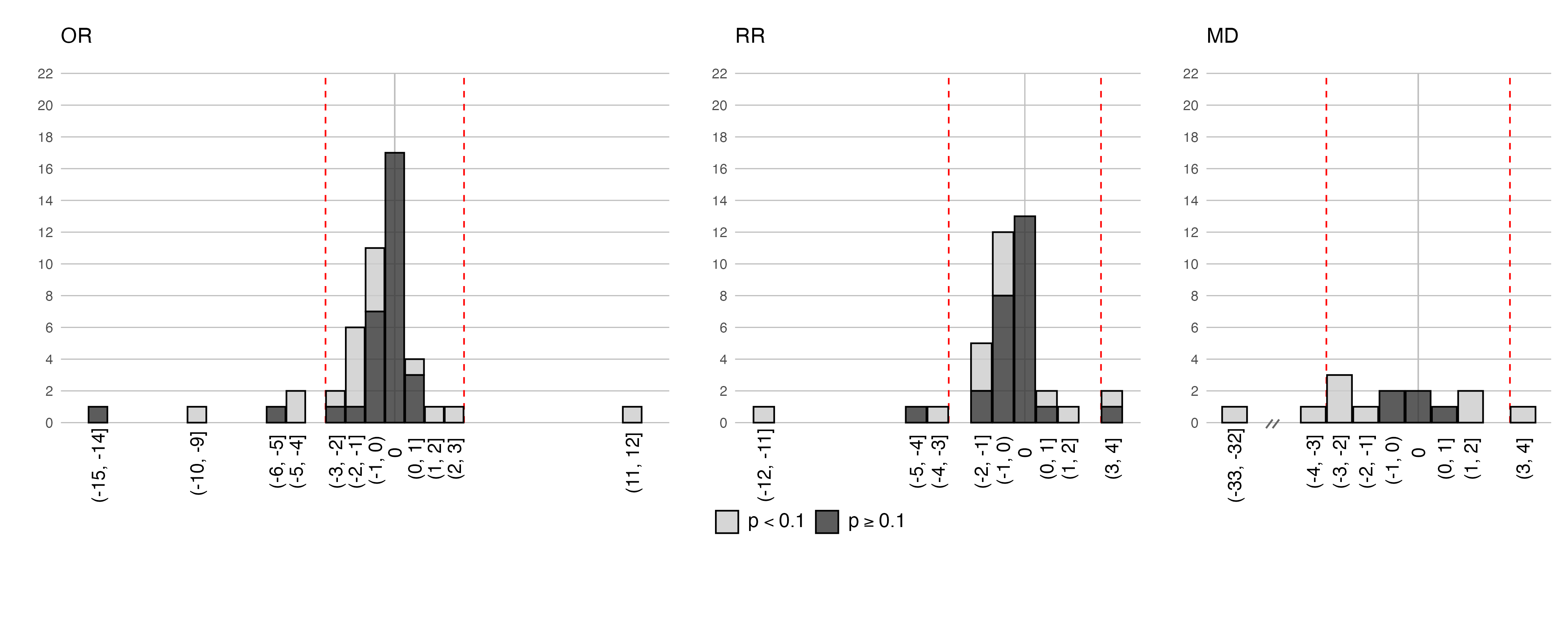}
 \caption{Distribution of $\Delta \aic = \aic\m - \aic\re$ for two-arm NMAs in the \texttt{nmadb} database, stratified by $p$-value for testing $H_0$ and effect measure (OR, RR, MD). Negative values favour the ME model, while positive values favour the RE model. Histogram bins generally span intervals of length one on the AIC scale, except for the special case $\Delta \aic = 0$, which is represented by a single‑value bin. Dashed lines at $\pm 3$ denote thresholds for meaningful differences in model fit. $\tau^2$ was estimated by the DL estimator. }
 \label{fig:AICdiff} 
\end{figure}

Figure~\ref{fig:AICdiff} presents histograms of $\Delta \aic$ for each effect measure, with red dashed lines at $\pm 3$ indicating the range within which the two models are generally considered to have comparable empirical support \citep{burnham2002model}. Datasets are further categorized by the significance of the heterogeneity test for $H_0$ ($p < 0.1$ v.s. $p \geq 0.1$). Quantiles of the distributions are also provided in Appendix~\ref{subsec_suppresults}, Table~\ref{tab:AIC_quantiles_DL}.

Overall, 86 of the 100 networks had $\Delta \aic \in [-3, 3]$, indicating that neither model provides a clearly superior fit in the majority of cases. Among these, 32 networks (17 OR, 13 RR, and 2 MD) had $\Delta\aic=0$ as a result of $\hat\phi=1$ and $\hat\tau^2_\text{DL}=0$ for the ME and RE models, respectively, implying they both reduce to the CE model. Among the 68 remaining networks with a preference for either model ($\Delta\aic \neq 0$ in Figure~\ref{fig:AICdiff}), a substantial proportion favoured the ME model, despite the RE model being the conventional choice in practice. The ME model yielded a lower AIC in 24/31 OR networks, 20/25 RR networks, and 8/12 MD networks. The respective median $\Delta\aic$ for each effect measures were negative: -0.51, -0.60, and -0.85. These findings suggest that, even though the ME model does not universally outperform the RE model, it provides equal or better fit in a substantial proportion of cases and warrants consideration alongside the conventional approach.

Among 31 OR, 26 RR and 5 MD networks without evidence of heterogeneity ($p\geq 0.1$), 6.45\%, 7.69\% and 0\% had $\Delta \aic \notin [-3, 3]$. In contrast, among 17 OR, 12 RR and 9 MD networks with evidence of heterogeneity ($p<0.1$), 23.5 \%, 25\% and 33.3\% had $\Delta \aic \notin [-3, 3]$, indicating that significant differences in the fit of the ME and RE model are more prevalent under evidence of heterogeneity. The corresponding median $\Delta \aic$s were -1.07, -0.55, and -2.02, respectively indicating a slight overall preference toward the ME model in most heterogeneous networks. Comparatively, the median $\Delta \aic$s were respectively 0 in networks with little evidence of heterogeneity ($p\geq0.1$), with the RE and ME models generally providing very similar empirical fit. %Additional quantiles are provided in (Appendix~\ref{subsec_suppresults} Table~\ref{tab:AIC_quantiles_DL}).

Ten networks exhibited a clear preference for the ME model  ($\Delta\aic<-3$), while four favoured the RE model ($\Delta\aic<-3$). Among these, five networks showed particularly pronounced differences ($|\Delta \aic| < -9$) and are therefore examined in greater detail as case studies in Section \ref{subsec_casestudies}.

\subsection{Case Studies}
\label{subsec_casestudies}

Three networks with $\Delta \aic < -9$ and $p < 0.1$ are presented as case studies in Sections~\ref{subsubsec_cs2}–\ref{subsubsec_cs3}. An additional network exhibiting $\Delta \aic = -32.6$ due to incorrect reporting of standard errors is presented in Appendix~\ref{subsubsec_cs1}. A single network with $\Delta\aic>9$ and $p < 0.1$ was investigated as case study in Section~\ref{subsubsec_cs4}. Finally, one network with $\Delta \aic < -9$ but $p \geq 0.1$ is investigated in Section~\ref{subsubsec_cs5}.

\subsubsection{Topical NSAIDs Versus Placebo for Pain Relief}
\label{subsubsec_cs2}

\begin{figure}[htbp]
\centering
\includegraphics[width=0.9\textwidth]{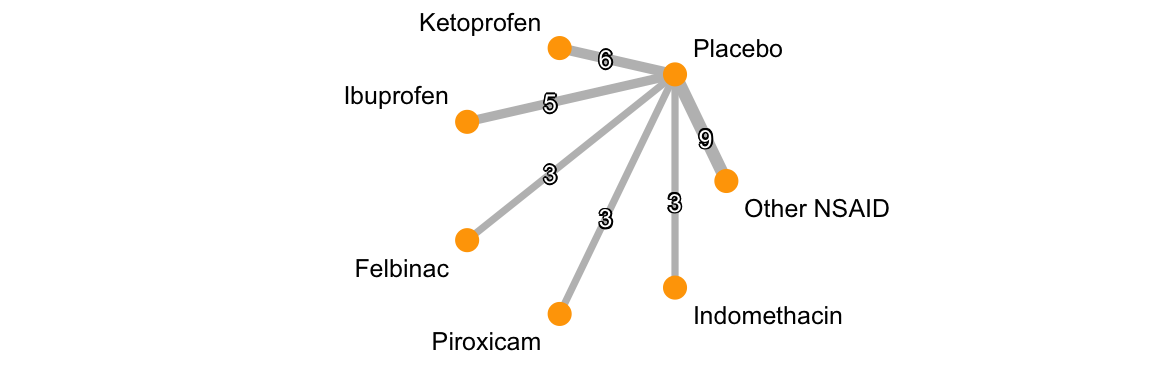}
\caption{Network of evidence on topical nonsteroidal anti-infammatory drugs (NSAIDs) for pain relief. Nodes represent interventions; edge widths are proportional to the number of direct comparisons.}
\label{fig:netgraph_1}
\end{figure}

\begin{figure}[htbp]
\centering
\includegraphics[width=\textwidth]{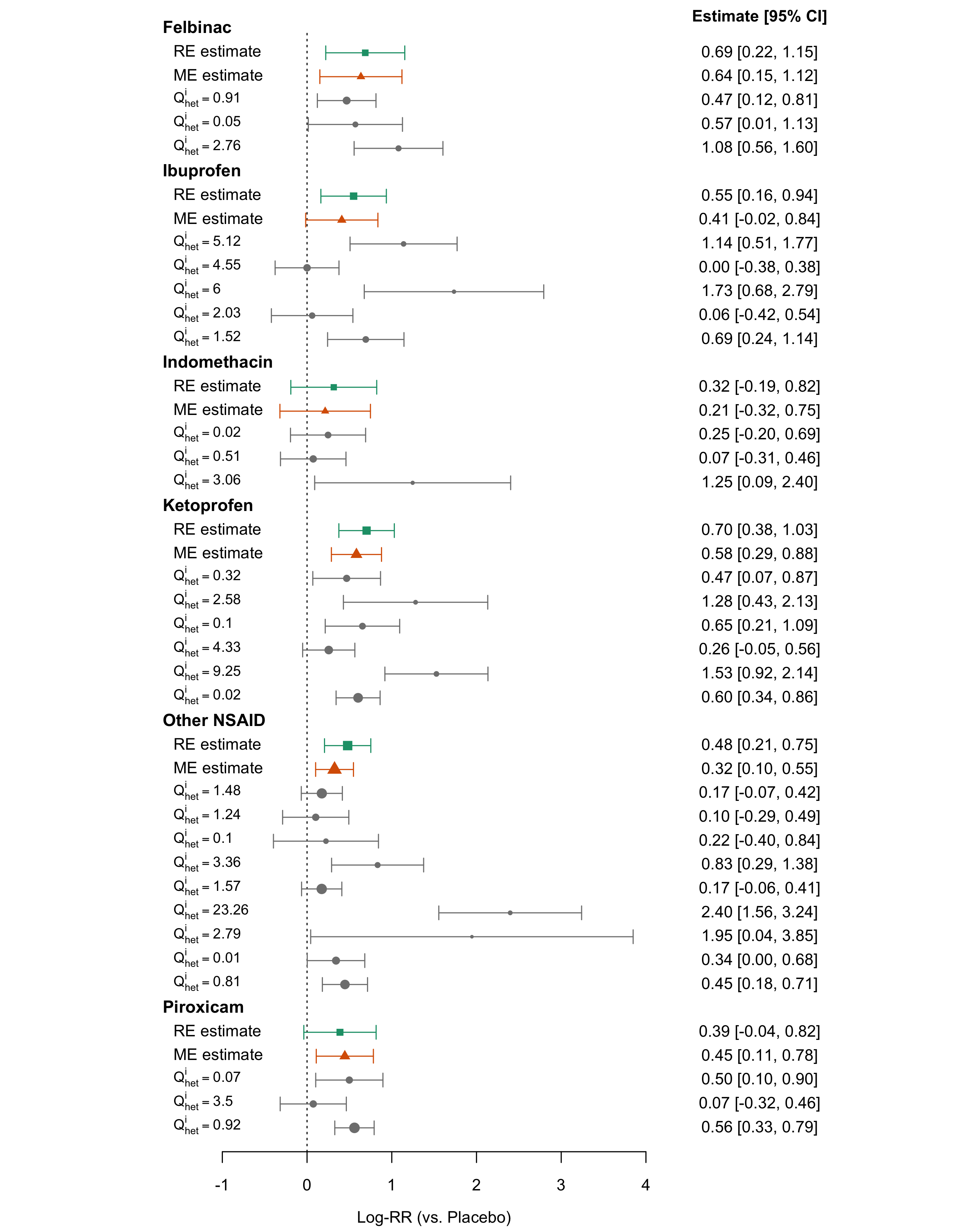}
\caption{Forest plot of log-risk ratio versus control for case study 1. Grey circles indicate each study’s estimate, with an area proportional to the study's weight (1/sample variance); the intervals represent the corresponding 95\% CIs. Numeric labels on the left provide study‐level heterogeneity contributions $\qh^i$. Green squares and orange triangles denote the overall RE and ME estimates, respectively. }
\label{fig:forest_1}
\end{figure}

In an NMA of $29$ studies comparing seven classes of topical nonsteroidal anti-inflammatory drugs (NSAIDs) versus placebo for achieving at least 50\% pain relief in adults with acute pain \citep[Figure \ref{fig:netgraph_1}, Appendix \ref{appen_q} Table \ref{tab:het_decomp1}]{mason2004topical}, the ME model significantly outperformed the additive RE model ($\Delta\aic =  -11.84$). The original synthesis by \citet{mason2004topical} consisted of six CE pairwise MAs, meaning no NMA was conducted in the original publication. We fitted both RE and ME NMA models using the same data and log-RR as the effect measure.

To aid identification of outlying studies contributing heterogeneity and their impact on model fit, a forest plot in Figure \ref{fig:forest_1} displays pooled estimates of the relative effects versus placebo from both ME and RE models, alongside individual study's contribution to heterogeneity $\qh^i$. 

The network exhibits significant heterogeneity, with $\qt = \qh = 82.25$  ($p$ = $1.37 \times 10^{-8}$), and $\qi = 0$ because it is a star network. A breakdown of $Q_\text{het}$ reveals that the other NSAID-placebo contrast ($\qh^c= 34.62$) and the ibuprofen-placebo contrast ($\qh^c= 19.22$) contribute most to heterogeneity. Smaller contributions come from the ketoprofen-placebo contrast ($\qh^c=16.6$) and the remaining contrasts vs. placebo.

Notably, one other NSAID-placebo study contributes $\qh^i=23.26$ \citep{Predel2004}, accounting for over one quarter of the total heterogeneity. This study evaluated a novel NSAID, diclofenac patch, which is likely to differ in efficacy compared to the 8 distinct NSAIDS lumped in the other NSAID category. As such, the heterogeneity in the other NSAID-placebo comparison is not surprising. Exclusion of this single outlier and refitting both models yields $\Delta\aic=-6.53$ and $\qh = 58.53$ ($p$ = $3.676179\times 10^{-5}$), demonstrating a reduction in between-study variability, yet still a clear preference for the ME model.

The ibuprofen-placebo contrast was the second largest source of heterogeneity in the network based on the sum of $\qh^i$. Five out of six studies contributing direct evidence to this comparison evaluated 5\% ibuprofen cream or gel with effect measure assessed after 7 days, while the other study \citep{Baracchi1982} investigated 10\% ibuprofen cream and assessed reduction in pain after 14 days, yielding the largest log-RR. Differences in drug doses and effect measure definitions could explain some of the heterogeneity. Moreover, even among the five studies using 5\% ibuprofen, the source of the patients' pain varied (ankle joint, ankle sprain, acute tendinitis, soft tissue injury, soft tissue trauma). Given the small number of studies, this heterogeneity would be difficult to mitigate in a network meta-regression model. As such, RE or ME models may be more appropriate to account for this heterogeneity. 

For the ketoprofen-placebo contrast, one study \citep{Noret1987} produces the second largest individual contribution to heterogeneity ($\qh^i = 9.25$). While this study reported the largest effect, its ketoprofen dose (2.5\% gel) is not the strongest among the included trials. Apart from differences in the source of the patients' pain, no other notable differences in the characteristics of the studies were noted, emphasizing that heterogeneity cannot always be explained and hence mitigated.

Overall, the pooled estimates from the ME model appear less sensitive to extreme study-level effects compared to those from the RE model (Figure~\ref{fig:forest_1}). This behaviour is consistent with the ME model’s weighting structure, which down-weights imprecise studies more than the additive RE model. Consequently, in this network, characterized by marked and unevenly distributed heterogeneity, the ME model provides a better empirical fit and more robust pooled estimates.

\subsubsection{Interventions to Increase Household Possession of a Functioning Smoke Alarm}\label{subsubsec_cs3}

\begin{figure}[htbp]
\centering
\includegraphics[width=0.9\textwidth]{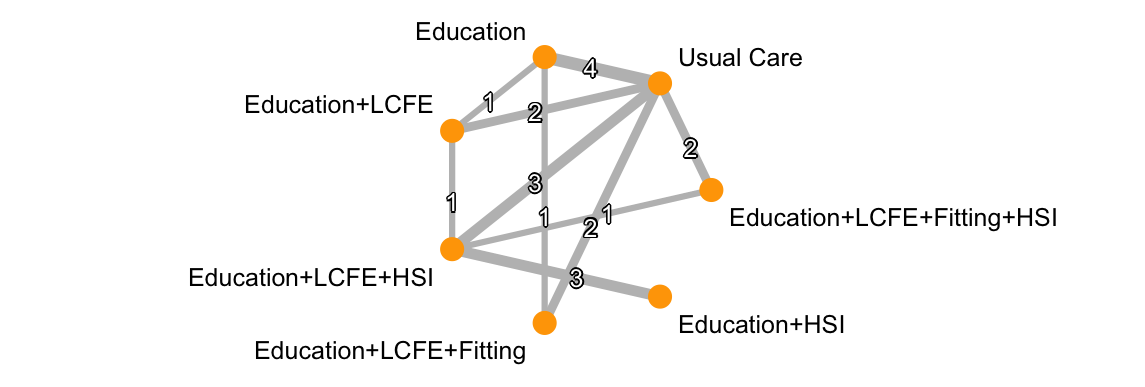}
\caption{Network of evidence on interventions for increasing household possession of functioning smoke alarms. Nodes represent interventions; edge widths are proportional to the number of direct comparisons. Abbreviations: LCFE stands for low-cost/free equipment, HSI stands for home safety inspection.}
\label{fig:netgraph_2}
\end{figure}

\begin{figure}[htbp]
\centering
\includegraphics[width=\textwidth]{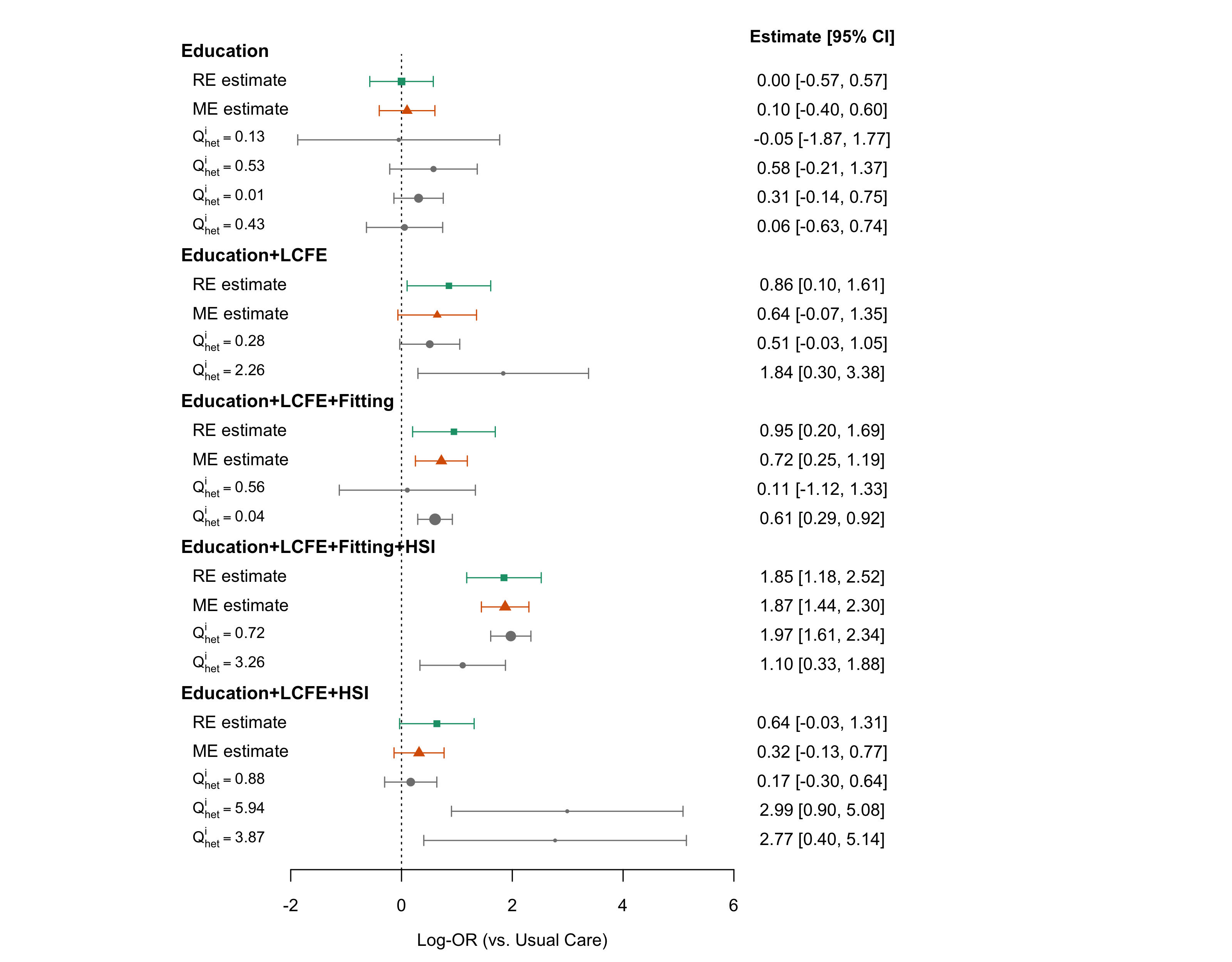}
\caption{Forest plot of log-odds ratio versus control for case study 2. Grey circles indicate each study’s estimate, with an area proportional to the study's weight (1/sample variance); the intervals represent the corresponding 95\% CIs. Numeric labels on the left provide study‐level heterogeneity contributions $\qh^i$. Green squares and orange triangles denote the overall RE and ME estimates, respectively.}
\label{fig:forest_2}
\end{figure}

In an NMA of $20$ studies comparing seven interventions to improve household possession of a functioning smoke alarm \citep[Figure~\ref{fig:netgraph_2}, Appendix \ref{appen_q} Table \ref{tab:het_decomp2}]{cooper2012network}, the interventions were compared on the log-odds ratio scale. The ME model was preferred over the additive RE model, with $\Delta\aic =  -9.02$. 

Total heterogeneity was $\qt = 36.12$ ($p$ = 0.001), indicating significant between‐study variation.  A decomposition of $\qh=23.51$ ($p=0.009$) revealed that the usual care versus Education+LCFE+HSI contrast contributed the most to heterogeneity ($\qh^c=10.7$), driven largely by two RCTs contributing $\qh^i=3.87$ and $5.94$ \citep{hendrickson2002safety,sangvai2007studying}. These two studies reported larger effects and had larger standard errors than the third study contributing direct evidence, which was not a RCT. Differences in the study-specific definitions of usual care could be a possible driver of this heterogeneity, but this was not examined closely in the original NMA.

Another influential study came from the Education+LCFE+HSI vs. Education+HSI contrast: one study contributed $\qh^i=4.53$ and reported a considerably large log‐OR of $-3.15$ ($s=1.46$) \citep{johnston2000preschool}, while the other two studies comparing the same treatments reported log‐ORs of $0.0000$ ($s_i = 0.82$) and $0.0168$ ($s_i = 0.18$). Limited information in the original NMA was available to explain the discrepancy, though any differences in the study-specific definitions of the non-pharmacological interventions (e.g., duration of education) could be a factor. 

This network also exhibited evidence of inconsistency ($\qi = 12.61$, $p=0.01$), suggesting that the lack of fit of a CE model in this network is not driven solely by within-design heterogeneity, but also by disagreement between direct and indirect evidence. The ME model’s lower AIC may reflect a greater flexibility in accommodating inconsistency through multiplicative heterogeneity in this network. However this should be interpreted cautiously, as inconsistency may not be fully accommodated. To examine this, an unrelated mean effects ME model \citep{dias2013inconsistency} was fitted and showed an AIC 2.14 higher than the consistency ME model, alleviating concern about residual inconsistency.% The AIC under UME model was 61.12, which is larger than that under the ME model ($\aic\m = 50.81$). Thus, despite evidence of inconsistency from the $Q$-statistic, the consistency ME model remained preferred by AIC in this network. 

Of 20 studies included in the NMA by \citet{cooper2012network}, 14 were RCTs, and 6 were non-RCTs. Should the RCTs be deemed more trustworthy than the non-RCTs, then the NMA should have been run without the non-RCTs at least as a sensitivity analysis. When we refitted the CE model using only RCTs, we found no detectable heterogeneity within designs but still found evidence of inconsistency ($\qt = 32.98$, $p=6.2\times10^{-5}$, $\qi = 27.62$, $p=1.5\times10^{-5}$, $\qh = 5.35$, $p$ = 0.25).

\subsubsection{Biological Therapies Versus Placebo or Standard Care for ACR70 Improvement}

\label{subsubsec_cs4}
\begin{figure}[htbp]
\centering
\includegraphics[width=0.9\textwidth]{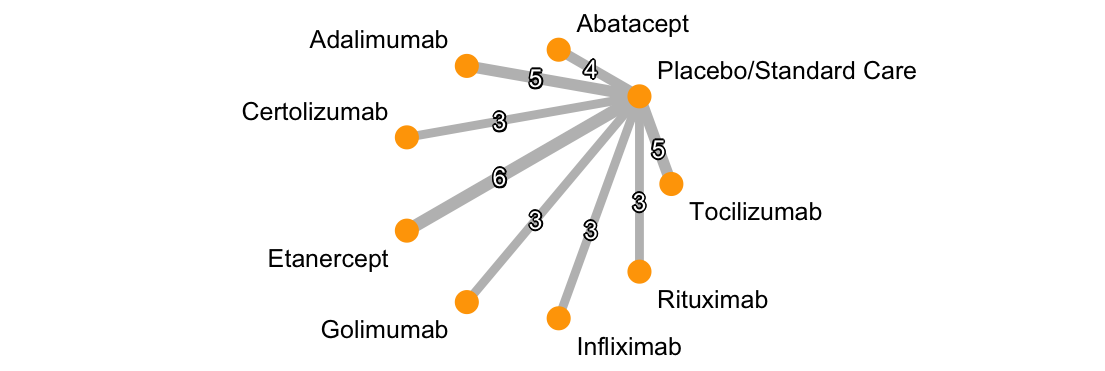}
\caption{Network of evidence on biologic treatments for ACR70 improvement. Nodes represent interventions; edge widths are proportional to the number of direct comparisons.}
\label{fig:netgraph_3}
\end{figure}

\begin{figure}[htbp]
\centering
\includegraphics[width=\textwidth]{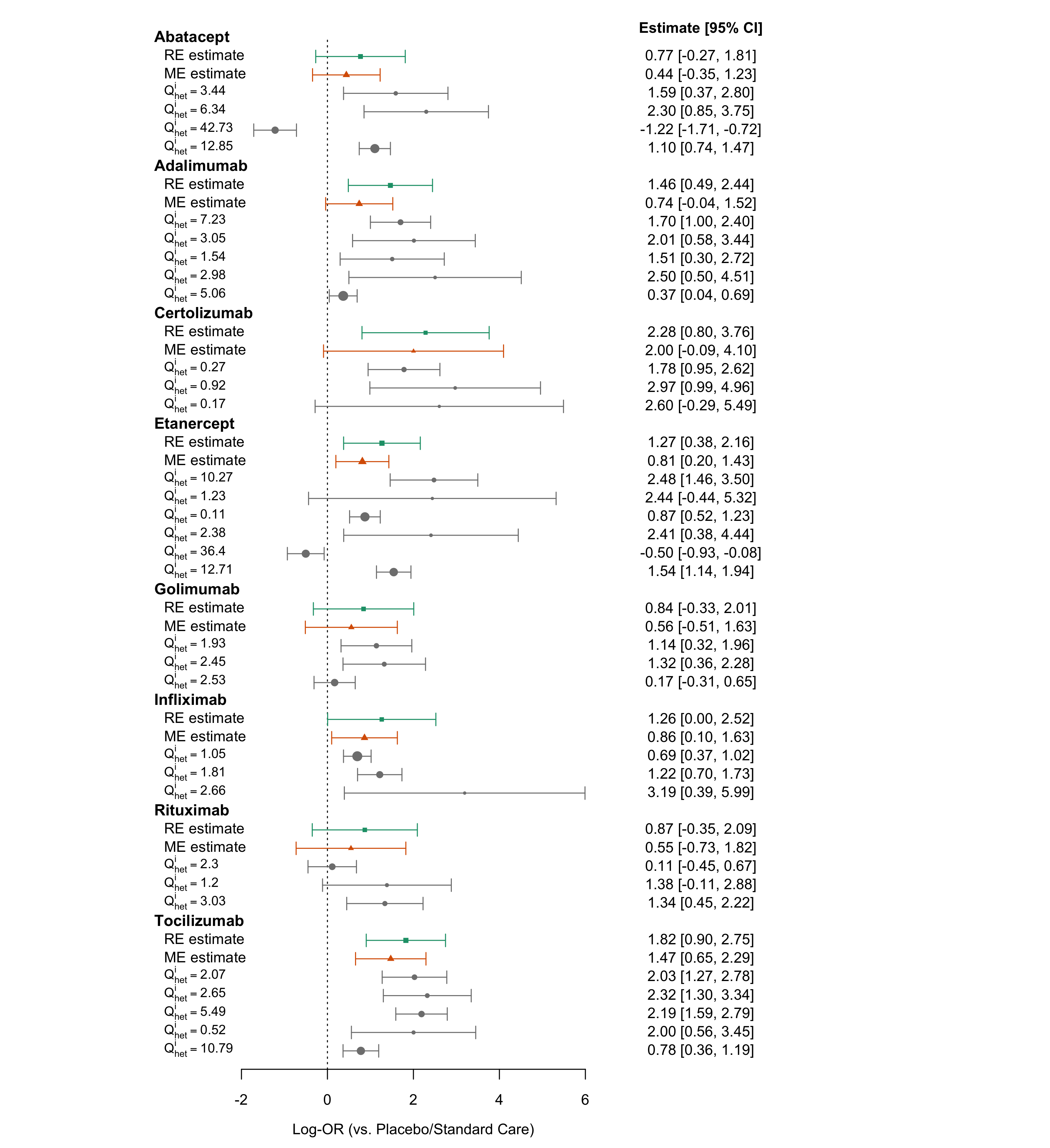}
\caption{Forest plot of log-odds ratio versus control for case study 3. Grey circles indicate each study’s estimate, with an area proportional to the study's weight (1/sample variance); the intervals represent the corresponding 95\% CIs. Numeric labels on the left provide study‐level heterogeneity contributions $\qh^i$. Green squares and orange triangles denote the overall RE and ME estimates, respectively.}
\label{fig:forest_3}
\end{figure}

An NMA of 32 studies comparing eight biological therapies against placebo or standard care for 70\% improvement in the American College of Rheumatology criteria (ACR70) \citep[Figure~\ref{fig:netgraph_3}, Appendix \ref{appen_q} Table \ref{tab:het_decomp3}]{brodszky2011rheumatoid} was previously synthesized through a RE network meta-regression (NMR) model on the log-OR scale, adjusting for effect modifiers such as methotrexate use, disease duration, and baseline HAQ score. As these effect modifiers were not available in \texttt{nmadb}, both RE and ME models were fitted without adjustment. The additive RE model was strongly preferred over the ME model with $\Delta\aic =  11.47$. 

Heterogeneity was significant ($\qt = \qh = 190.15$, $p$ = $8.31 \times 10^{-28}$). As shown in Figure \ref{fig:forest_3}, we observed some outliers with relatively narrower CIs; as the standard error is smaller in these studies, they are given more weight in the ME model compared to the RE model. In particular, two studies have negative log-odds ratios, implying these treatments are worse than placebo in improving ACR70, and contributed $\qh^i=42.73$ (abatacept vs placebo, log-OR = -1.22) \citep{bathon2000comparison} and $\qh^i=36.40$ (etancercept vs placebo, log-OR = -0.504) \citep{kremer2006effects} to total heterogeneity. Both studies favoured placebo over the active treatment, which conflicted with what was observed in other studies making the same comparison. 

The study by \citet{bathon2000comparison} in the abatacept-placebo contrast enrolled patients with notably shorter disease duration (1 year) compared to the other three studies making the same comparison. In addition, methotrexate was only given to patients in the placebo group; the other studies either assigned methotrexate to both treatment groups or neither treatment group. For the etanercept-placebo contrast, \citet{kremer2006effects} recruited methotrexate-resistant patients, a population not represented in other etanercept-placebo trials. These design and population differences plausibly explain the discordant results.

Refitting both models after excluding these two outlying studies reduced total heterogeneity ($\qt=78.37$, $p$= $2.99\times 10^{-8}$), but preserved a preference for the RE model ($\Delta \aic = 5.88$). Overall, in this network, the ME model appeared to be more sensitive to extreme but precise observed effects.

\subsubsection{Novel Oral Anticoagulants for Atrial Fibrillation}
\label{subsubsec_cs5}

\begin{figure}[htbp]
\centering
\includegraphics[width=0.9\textwidth]{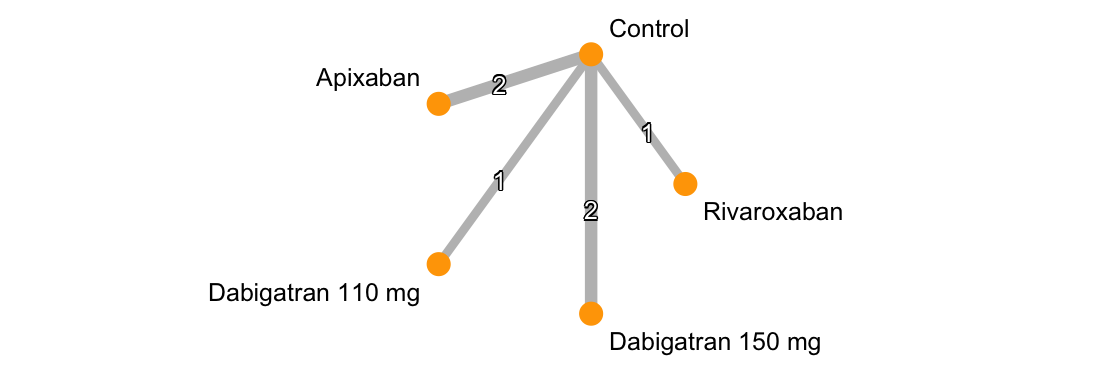}
\caption{Network of evidence on novel oral anticoagulants for atrial fibrillation. Nodes represent interventions; edge widths are proportional to the number of direct comparisons.}
\label{fig:netgraph_4}
\end{figure}

\begin{figure}[htbp]
\centering
\includegraphics[width=\textwidth]{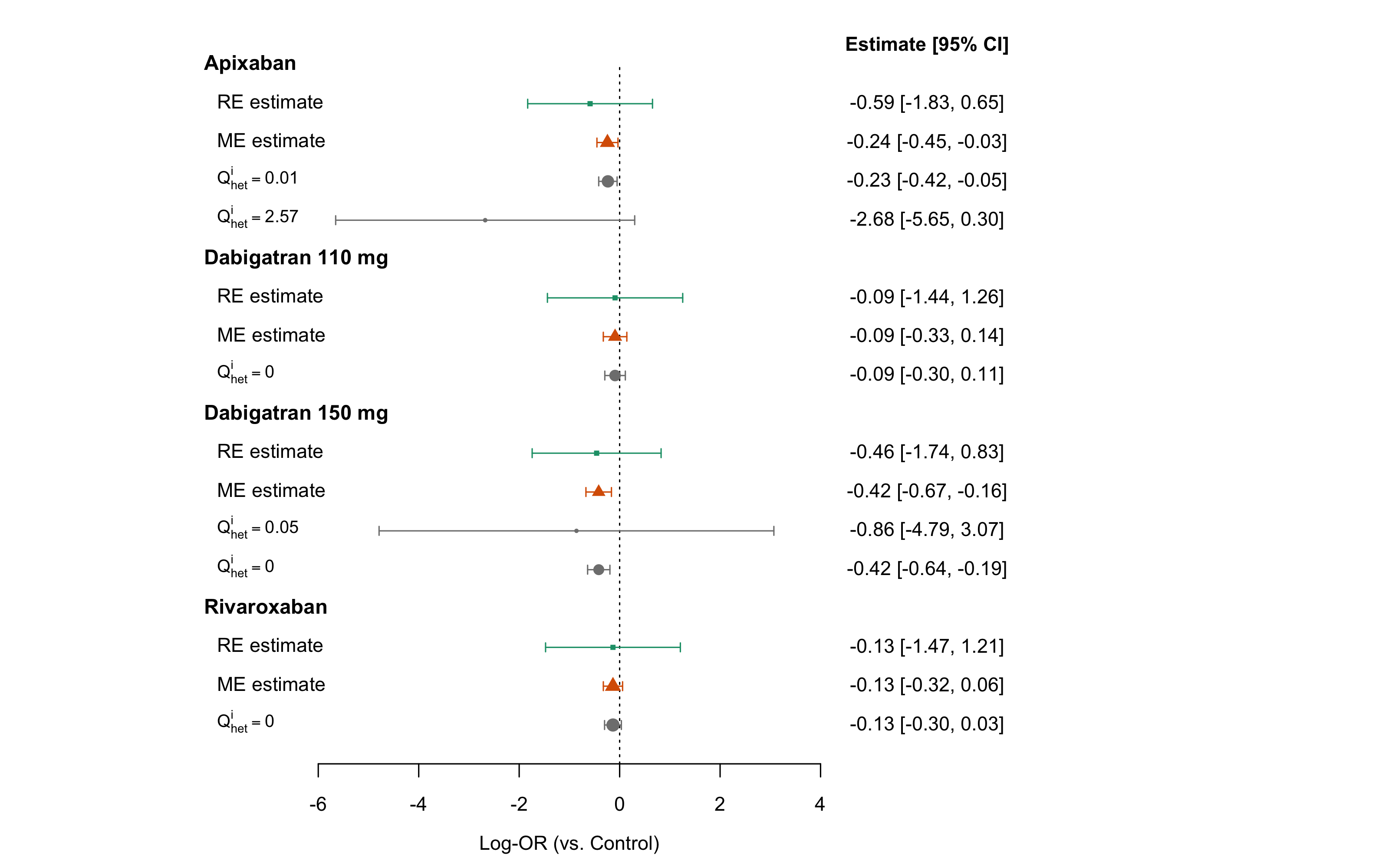}
\caption{Forest plot of log-odds ratio versus control for case study 4. Grey circles indicate each study’s estimate, with an area proportional to the study's weight (1/sample variance); the intervals represent the corresponding 95\% CIs. Numeric labels on the left provide study‐level heterogeneity contributions $\qh^i$. Green squares and orange triangles denote the overall RE and ME estimates, respectively.}
\label{fig:forest_4}
\end{figure}

This NMA examines four novel oral anticoagulants versus control on the outcome of long-term stroke or systemic embolism \citep[Figure~\ref{fig:netgraph_4}, Appendix \ref{appen_q} Table \ref{tab:het_decomp4}]{biondi2013comparative}. Model comparison based on the AIC strongly favours the ME model over the RE model, with $\aic\m - \aic\re = -14.07$. The overall level of heterogeneity is not statistically significant, with $\qt = \qh = 2.63$ ($p = 0.27$). The estimated multiplicative heterogeneity parameter is $\hat{\phi} = 1.31$, and the additive between-study variance is $\hat{\tau}^2 = 0.46$. 

The original study by \citet{biondi2013comparative} also conducted pairwise meta-analyses and assessed small-study effects via funnel plot inspection. Evidence of small-study effects was observed, primarily attributable to the inclusion of small phase II trials. In particular, one study comparing apixaban with control \citep{ogawa2011safety}, with a sample size of $n = 222$, contributed disproportionately to the heterogeneity, with $\qh^i = 2.57$ ($p = 0.11$). In contrast, another study evaluating the same comparison \citep{granger2011apixaban} included $n = 18{,}201$ participants and therefore provided much more precise estimates. 

This imbalance highlights the sensitivity of the RE model to small, potentially outlying studies. The RE model assigns relatively greater weight to smaller studies compared to the ME model, and provides more conservative CIs in this network. As a result, the presence of small-study effects can lead to inflated uncertainty and wider CIs for treatment effects under the RE model. In contrast, the ME model accommodates such variability through a multiplicative scaling of the within-study variances, which appears more appropriate in this setting given the absence of strong evidence for additive heterogeneity.

%A pairwise RE meta-analysis of the two trials comparing apixaban with control gives $\hat\tau = 1.35$ and an estimated treatment effect of $-0.98$ with 95\% confidence interval $[-3.19, 1.22]$, whereas the corresponding NMA RE estimate is $-0.59$ with 95\% confidence interval $[-1.83, 0.65]$. 

% AIC_RE - AIC_CE = 15.1, AIC_ME-AIC_CE = 1.01

\section{Discussion} \label{s:discussion}

We conducted an empirical comparison of the conventional additive random-effects (RE) model and the multiplicative-effect (ME) model across 100 NMAs of two-arm studies summarizing odds ratios, risk ratios, and mean differences extracted from the \texttt{nmadb} database. We quantified model fit using $\Delta\aic = \aic\m-\aic\re$. Better or equal fit of the ME model ($\Delta\aic \leq 0$) compared to an RE model fitted via the DL method was observed in 84 out of 100 NMAs, which aligns with the 79.4\% figure found by \cite{stanley2023unrestricted} in an evaluation of 67,308 Cochrane MAs. A significant difference between models, indicated by $|\Delta\aic|> 3$, was found in 10 of 38 networks (26 \%) with significant heterogeneity ($p$ < 0.1) but only in 4 of 62 networks (6 \%) without such evidence ($p\geq$  0.1). The ME model was substantially preferred ($\Delta\aic<-3$) in ten NMAs, while the RE model was substantially preferred ($\Delta\aic>3$) in four NMAs. These results demonstrate that the ME formulation often provides equal or better fit, and therefore merits consideration besides the conventional RE model.

Case studies 1, 2, 3, and 4 (Sections \ref{subsubsec_cs2}–\ref{subsubsec_cs4}) reveal how variance patterns drive model preference. In both case study 1 (topical NSAIDs) and case study 2 (smoke‐alarm interventions), total heterogeneity was driven largely by the less precise studies. The ME model’s variance inflation factor $\phi$ allowed for better down‐weighting of imprecise trials, yielding lower $\aic\m$ than $\aic\re$. Similarly, case study 4 (novel oral anticoagulants for long-term stroke or systemic embolism) showed a strong preference for the ME model despite non-significant heterogeneity, likely because variability was driven by a small, imprecise phase II trial that received relatively greater weight under the RE model. Conversely, in case study 3 (biological therapies for ACR70), most heterogeneity arose from precise studies with extreme treatment effects. The RE model assigns less weight to these outliers compared to the ME model, but context experts would have to be consulted to assess whether either weighting scheme reflects the trustworthiness of the estimates. Importantly, observed differences in model fit should not be attributed solely to variance structure. Unaccounted effect modifiers can induce heterogeneity, potentially influencing the preference between ME and RE models.

Several insights arise from our analyses. First, the ME model is more supported by the data when heterogeneity arises from imprecise studies. Second, individual outlying studies can influence the total heterogeneity $\qh$ by a large amount; accordingly, sensitivity checks are essential. Third, although the ME model down‐weights high‐variance studies, the resulting confidence interval width depends jointly on the estimated scaling factor $\hat{\phi}$ and the within‐study standard errors. Consequently, the ME model may yield wider or narrower intervals than the RE model depending on the network structure. Notably, the ME model produces identical point estimates for the treatment effects as the CE model, while adjusting the CIs by the multiplicative factor $\hat\phi$. This property preserves the interpretability and computational simplicity of the CE estimator, yet allows flexible accommodation of between‐study heterogeneity. Finally, our findings suggest that heterogeneity may vary across pairwise comparisons within the same network, indicating that a single global heterogeneity parameter may not always adequately capture the complexity of real‐world evidence networks \citep{lu2009modeling}.

Our study focuses on NMAs composed exclusively of two-arm trials. Such designs are common and provide a comparatively simpler setting for assessing heterogeneity than multi-arm trials, where a single heterogeneity parameter --whether in an RE or ME model -- may be insufficient to fully capture both heterogeneity and inconsistency. However, this focus on two-arm trials limits the generalizability of our findings, and extending the analysis to more complex network structures will require consideration of more flexible modelling frameworks. Another limitation is that our empirical evaluation relies on the \texttt{nmadb} package, whose curated networks span only 1999 to 2015. An older evidence base may not reflect the methodologies and characteristics of more recent networks. This also limits our ability to conduct extensive empirical evaluations using current best practices to minimize and explain heterogeneity (e.g., not lumping treatments into groups, accounting for known effect modifiers through regression). Future research should address these limitations in several directions. Updating the empirical evaluation to include NMAs with more recent trials will ensure the findings are relevant to current clinical practice and methodological advancements. Future empirical studies may assess the Deviance Information Criterion in a Bayesian setting when comparing RE and ME models. Such work could further illustrate the comparative strengths and limitations of the ME and RE models under a wider range of scenarios, providing deeper insights into their performance in various contexts.

In summary, our empirical analyses indicate that the ME model is more influenced by extreme and precise estimates, whereas the RE model is more influenced by extreme and imprecise estimates. This distinction implies that the ME formulation in NMA may offer greater resilience to small-study and related publication bias, where small-studies reporting extremely large effects tend to be published more than those with less extreme estimates; such small studies are often considered not trustworthy due to their increased susceptibility to chance fluctuations \citep{sterne2000publication}. While the reduced sensitivity of ME model to small, imprecise studies parallels insights from findings in pairwise MA \citep{stanley2023unrestricted}, our results demonstrate how this behaviour extends to NMA and depends critically on the interaction between study precision and extremeness of effects.
\end{spacing}
%TC:ignore

\section*{Acknowledgements}

This research was supported by A. Béliveau's NSERC Discovery Grant (RGPIN-2019-04404).

\section*{Conflict of interest statement}

The authors declare that they have no conflicts of interest.

\section*{Data availability statement}

The data that supports the findings of this study are available in the supplementary material of this article.

\clearpage

\bibliographystyle{plainnat}
\bibliography{ref}  

\setcounter{page}{1}
\setcounter{section}{0}
\setcounter{table}{0}
\setcounter{figure}{0}
\setcounter{equation}{0}

\makeatletter
\renewcommand \thesection{S\@arabic\c@section}
\renewcommand \thetable{S\@arabic\c@table}
\renewcommand \thefigure{S\@arabic\c@figure}
\renewcommand \theequation{S\@arabic\c@equation}
\makeatother

\newpage
\appendix
\section{Appendix}
\label{appen}

\subsection{Explanation of Exclusion of Multi-arm Trials}
\label{appen_multi_arm}
Our study focuses on NMAs composed exclusively of two-arm trials, which are common in practice and allow for a clear assessment of heterogeneity within simple network structures. We also reviewed an additional 140 NMAs from the \texttt{nmadb} database that included multi-arm trials; however, many of these datasets exhibited features that complicate the interpretation of the RE and ME model comparisons. In particular, several networks showed substantial variation in heterogeneity across treatment comparisons, suggesting that the assumption of a single common heterogeneity parameter may be inappropriate. Given that multi-arm trials are internally consistent by design, this pattern further raised concerns that inconsistency in other parts of the network may be more difficult to accommodate through ME or RE consistency models because they accommodate both heterogeneity and inconsistency through a single heterogeneity parameter. Under these circumstances, differences in $\Delta\aic$ may reflect broader model misspecification rather than meaningful preference between the heterogeneity formulations. Consequently, we chose to present results only for NMAs composed of two-arm trials. While including networks with multi-arm studies would provide a more comprehensive empirical evaluation, further methodological development is needed to adequately address the additional complexities commonly encountered in these datasets.

\subsection{Relationship between $\hat{\phi} = 1$ and $\hat{\tau}_\text{DL}^2=0$}
\label{appen_proof}

We aim to show that $\hat{\phi} = 1$ if and only if $\hat{\tau}^2_{\text{DL}} = 0$. Note that the leverages $h_i$ appearing in $\hat{\tau}^2_{\text{DL}}$, defined as $h_{i} = [\mathbf X(\mathbf X^\top\mathbf V^{-1}\mathbf X)^{-1}\mathbf X^\top\mathbf V^{-1}]_{ii}$, satisfy $0 \le h_i \le 1$.

We first consider the special case in which $h_i = 1$ for all $i$, so that all fitted values coincide with the observed data, implying $Q_{\text{total}} = 0$. An example of this setting is a star network with one study per edge. In this case, the denominator in the expression $\hat{\tau}^2_{\text{DL}} = \max\left\{0,\frac{\qt - df_\text{total}}{\sum_{i=1}^{m}(1-h_{i})/s_i^2}\right\}$ is zero, so the estimator is not defined. However, since the data provide no information about heterogeneity, it is customary to set $\hat{\tau}^2_{\text{DL}} = 0$.

Similarly, if the denominator in the definition of $\hat{\phi} = \max\left\{1, \frac{\qt}{df_\text{total}} \right\}$ is zero, i.e., $df_{\text{total}} = 0$ (as in the star network described), we analogously would set $\hat{\phi} = 1$. If $df_{\text{total}} > 0$ but $Q_{\text{total}} = 0$ is realized, we obtain $\hat{\phi} = \max(1, 0) = 1$ as well.

Excluding the special case considered above, we assume that at least one $h_i \neq 1$, so that not all fitted values coincide with the observed data. Under this assumption, it follows that
$$
\frac{Q_{\text{total}}}{df_{\text{total}}} > 1 
\;\Longleftrightarrow\;
\frac{Q_{\text{total}} - df_{\text{total}}}{\sum_{i=1}^m (1 - h_i)/s_i^2} > 0,
$$
which establishes the result $\hat{\phi} > 1$ if and only if $\hat{\tau}^2_{\text{DL}} > 0$. The contrapositive follows directly.

\subsection{Quantile table of $\Delta\aic$ under DL method}
\label{subsec_suppresults_DL}
\begin{table}[htbp]
\centering
\caption{Summary of $\Delta \aic$ by effect measure, AIC configuration, and $p$-value for testing $H_0$. The RE model was fitted using the DL estimate for $\tau^2$.}
\label{tab:AIC_quantiles_DL}
\begin{tabular}{llrrrrrr}
  \hline
Effect measure & Inclusion criterion & Min & q25 & Median & q75 & Max & $n$ \\ 
  \hline
  Odds ratio & all & -14.07 & -1.04 & -0.00 & 0.00 & 11.47 &  48 \\ 
  Risk ratio & all & -11.84 & -0.88 & -0.32 & 0.00 & 3.97 &  38 \\ 
  Mean difference & all & -32.60 & -2.36 & -0.14 & 0.13 & 3.51 &  14 \\ 
  Odds ratio & $\Delta\aic\neq 0$ & -14.07 & -1.55 & -0.51 & -0.04 & 11.47 &  31 \\ 
  Risk ratio & $\Delta\aic\neq 0$ & -11.84 & -1.35 & -0.60 & -0.34 & 3.97 &  25 \\ 
  Mean difference & $\Delta\aic\neq 0$ & -32.60 & -2.54 & -0.85 & 0.58 & 3.51 &  12 \\ 
  Odds ratio & $p<0.1$ & -9.02 & -1.55 & -1.07 & -0.01 & 11.47 &  17 \\ 
  Risk ratio & $p<0.1$ & -11.84 & -1.51 & -0.55 & -0.27 & 3.91 &  12 \\ 
  Mean difference & $p<0.1$ & -32.60 & -2.72 & -2.02 & 1.82 & 3.51 &   9 \\ 
  Odds ratio & $p\geq0.1$ & -14.07 & -0.24 & 0.00 & 0.00 & 0.66 &  31 \\ 
  Risk ratio & $p\geq0.1$ & -4.90 & -0.73 & 0.00 & 0.00 & 3.97 &  26 \\ 
  Mean difference & $p\geq0.1$ & -0.17 & -0.10 & 0.00 & 0.00 & 0.17 &   5 \\ 
   \hline
\end{tabular}
\end{table}

Table~\ref{tab:AIC_quantiles_DL} summarizes the distribution of $\Delta\aic=\aic\m-\aic\re$ across NMAs stratified by effect measure, inclusion of networks with $\Delta\aic=0$, and significance of Cochran's $Q$ test for heterogeneity. Negative values favour the ME model, while positive values favour the additive RE model. Across all effect measures, the distributions are generally centred near zero, and the medians are negative except for $p\geq 0.1$, indicating that the ME model more often provides improved fit when differences between models occur. This tendency is more pronounced among heterogeneous networks ($p<0.1$), where the medians are uniformly negative. In contrast, non-heterogeneous networks ($p\geq0.1$) show medians and upper quartiles equal or close to zero, suggesting little practical distinction between the two models in the absence of detectable heterogeneity.

\subsection{Supplementary results for estimation by REML}
\label{subsec_suppresults}
\normalsize
\begin{figure}[H]
 \centering
 \includegraphics[width=\textwidth]{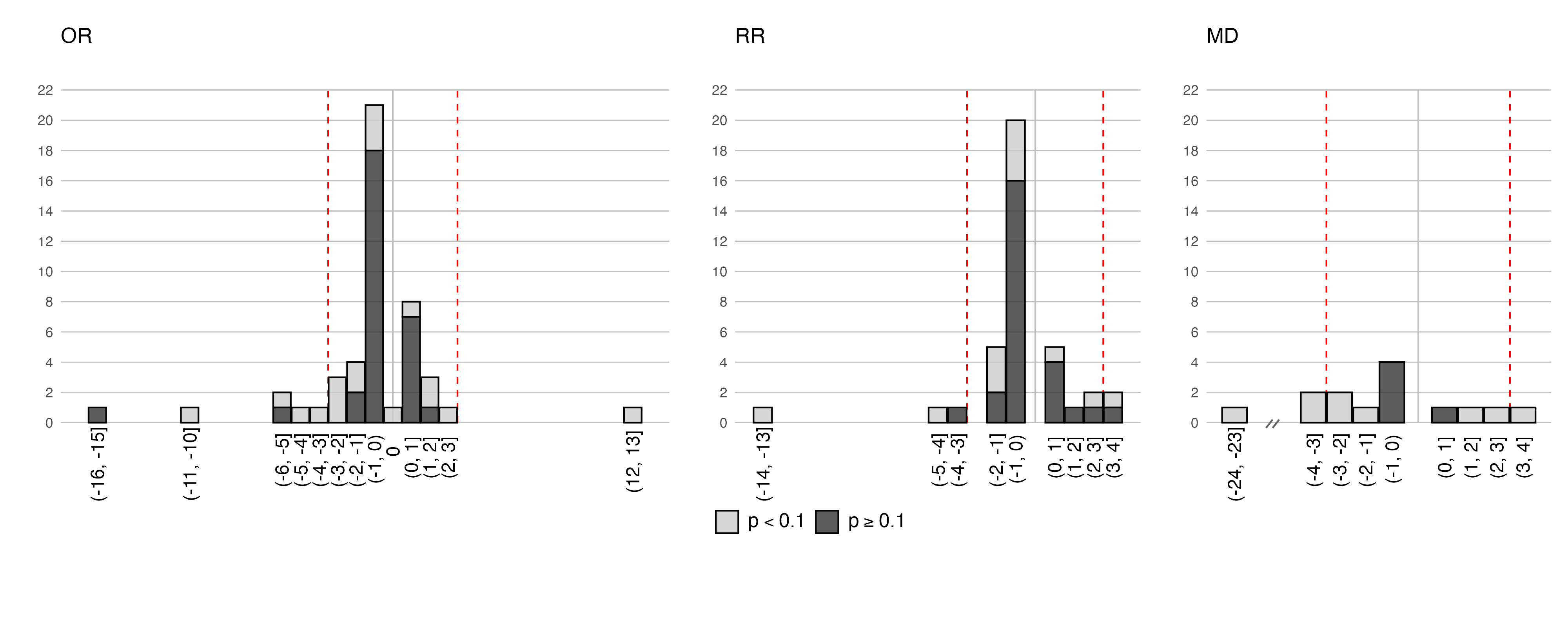}
 \caption{Distribution of $\Delta \aic = \aic\m - \aic\re$ for two-arm NMAs in the \texttt{nmadb} database, stratified by $p$-value of testing $H_0:\qt = 0$ and effect measure (OR, RR, MD). Negative values favour the ME model, while positive values favour the RE model. Histogram bins generally span intervals of length one on the AIC scale, except for the special case $\Delta \aic = 0$, which is represented by a single‑value bin. Dashed lines at $\pm 3$ denote thresholds for meaningful differences in model fit. $\tau^2$ estimated by REML.}
 \label{fig:AICdiff_REML}.
\end{figure}

\begin{table}[htbp]
\centering
\caption{Summary of $\Delta\aic$ by effect measure, AIC configuration, and $p$-value for testing $H_0$. The RE model was fitted using the REML estimator for $\tau^2$.}
\label{tab:aic_quantiles_REML}
\begin{tabular}{llrrrrrr}
  \hline
Effect measure & Inclusion criterion & Min & q25 & Median & q75 & Max & n \\ 
  \hline
Odds ratio & all & -15.32 & -1.04 & -0.00 & 0.02 & 12.53 &  48 \\ 
  Risk ratio & all & -13.16 & -0.70 & -0.00 & 0.04 & 3.83 &  38 \\ 
  Mean difference & all & -23.87 & -2.76 & -0.09 & 0.14 & 3.37 &  14 \\ 
  Odds ratio & $\Delta\aic\neq 0$ & -15.32 & -1.05 & -0.00 & 0.02 & 12.53 &  47 \\ 
  Risk ratio & $\Delta\aic\neq 0$ & -13.16 & -0.70 & -0.00 & 0.04 & 3.83 &  38 \\ 
  Mean difference & $\Delta\aic\neq 0$ & -23.87 & -2.76 & -0.09 & 0.14 & 3.37 &  14 \\ 
  Odds ratio & $p<0.1$ & -10.41 & -2.32 & -1.01 & 0.01 & 12.53 &  18 \\ 
  Risk ratio & $p<0.1$ & -13.16 & -1.29 & -0.70 & -0.22 & 3.83 &  12 \\ 
  Mean difference & $p<0.1$ & -23.87 & -3.25 & -2.61 & 1.95 & 3.37 &   9 \\ 
  Odds ratio & $p\geq0.1$ & -15.32 & -0.21 & -0.00 & 0.02 & 1.14 &  30 \\ 
  Risk ratio & $p\geq0.1$ & -3.43 & -0.35 & -0.00 & 0.06 & 3.17 &  26 \\ 
  Mean difference & $p\geq0.1$ & -0.12 & -0.06 & -0.00 & -0.00 & 0.19 &   5 \\ 
   \hline
\end{tabular}
\end{table}

Figure \ref{fig:AICdiff_REML} presents histograms of the $\Delta\aic$ distribution under REML estimation of $\tau^2$. In these histograms, $\Delta\aic = 0$ did not occur, except in one non-convergent sparse dataset where $\hat\tau^2 = 0$ was imposed. This contrasts with the DL approach, under which both the RE model and ME models are functions of the same heterogeneity statistic $\qt$, and $\hat\tau^2_{\text{DL}} = 0$ if and only if $\hat\phi = 1$ (see Appendix~\ref{appen_proof}), reducing both models to the CE model and yielding $\Delta\aic = 0$. In contrast, REML estimates $\tau^2$ through restricted likelihood maximization, which typically avoids boundary estimates at $\hat\tau^2 = 0$. As a result, $\Delta\aic = 0$ is essentially absent under REML estimation, although most values still concentrated within $\Delta\aic\in(-1,0]$.

Table~\ref{tab:aic_quantiles_REML} presents the same set of quantiles as Table~\ref{tab:AIC_quantiles_DL}, but with $\tau$ estimated via REML. Since only one NMA had $\Delta$AIC = 0, the results under the criterion $\Delta$AIC $\neq 0$ appear identical to those obtained from all 100 NMAs. Hence, the median values are not as strongly negative as those reported in Table~\ref{tab:AIC_quantiles_DL}.

\subsection{Rheumatoid Arthritis Treatments Versus Placebo for Pain Improvement}\label{subsubsec_cs1} 
An NMA comparing eight treatments for rheumatoid arthritis \citep{jansen2014comparative} yielded $\Delta\aic=-32.6$, indicating strong preference for the ME model. The effect measure of interest was pain improvement and data were synthesized as mean differences relative to placebo. Between-study heterogeneity is high ($\qt$ = 676.67  $p$ = $6.66 \times 10^{-143}$, with $\hat\phi = 112.78$ for the ME model and $\hat\tau_\text{DL} = 3.78$ for the RE model). 

We revisited the original dataset and discovered the standard errors provided by \texttt{nmadb} had been incorrectly divided by the square root of the sample size. For instance, in the study by \cite{van2004efficacy} comparing adalimumab (ADA) 40 mg to placebo, the change in pain from baseline on the 0 to 100mm visual analog scale (VAS) is -27.60 (standard error $= 2.93$, sample size = 113) for ADA and -11.00 (standard error $=2.55$, sample size = 110) for placebo. The standard error for the contrast ADA versus placebo should be calculated as $\sqrt{2.93^2+2.55^2}=3.88$, whereas in the \texttt{nmadb} database the standard errors were incorrectly divided by the sample sizes, giving a value of $\sqrt{2.93^2/113+2.55^2/110}=0.37$. Recomputing all standard errors correctly and refitting both ME and RE models yielded $\Delta\aic =-0.42$, indicating similar support for ME and RE models.

\subsection{Case Study Data and Q Decomposition}
\label{appen_q}
\begin{table}[H]
  \centering
  \caption{Study‐level contributions to $\qh$ and estimates of case study 1}
  \label{tab:het_decomp1}
  \begin{tabular}{llccc}
    \toprule
    Treatment 1         & Treatment 2         & $\qh^i$   & $y_i$  & $s_i$  \\
    \midrule
Placebo & Ketoprofen & 0.3210 & 0.4675 & 0.2039 \\ 
  Placebo & Ketoprofen & 2.5784 & 1.2809 & 0.4346 \\ 
  Placebo & Ketoprofen & 0.0999 & 0.6539 & 0.2242 \\ 
  Placebo & Ketoprofen & 4.3292 & 0.2561 & 0.1571 \\ 
  Placebo & Ketoprofen & 9.2509 & 1.5270 & 0.3103 \\ 
  Placebo & Ketoprofen & 0.0218 & 0.6026 & 0.1325 \\ 
  Placebo & Ibuprofen & 6.0049 & 1.7346 & 0.5405 \\ 
  Placebo & Ibuprofen & 2.0337 & 0.0601 & 0.2454 \\ 
  Placebo & Ibuprofen & 1.5152 & 0.6931 & 0.2300 \\ 
  Placebo & Ibuprofen & 5.1205 & 1.1394 & 0.3223 \\ 
  Placebo & Ibuprofen & 4.5503 & 0.0000 & 0.1922 \\ 
  Placebo & Felbinac & 0.9078 & 0.4675 & 0.1765 \\ 
  Placebo & Felbinac & 0.0528 & 0.5705 & 0.2836 \\ 
  Placebo & Felbinac & 2.7561 & 1.0797 & 0.2674 \\ 
  Placebo & Piroxicam & 0.0690 & 0.4990 & 0.2024 \\ 
  Placebo & Indomethacin & 0.0225 & 0.2478 & 0.2262 \\ 
  Placebo & Piroxicam & 3.5008 & 0.0733 & 0.1991 \\ 
  Placebo & Indomethacin & 0.5119 & 0.0728 & 0.1972 \\ 
  Placebo & Piroxicam & 0.9217 & 0.5596 & 0.1185 \\ 
  Placebo & Indomethacin & 3.0623 & 1.2465 & 0.5901 \\ 
  Placebo & Other NSAID & 1.4769 & 0.1743 & 0.1237 \\ 
  Placebo & Other NSAID & 1.2433 & 0.1023 & 0.1994 \\ 
  Placebo & Other NSAID & 0.1029 & 0.2231 & 0.3162 \\ 
  Placebo & Other NSAID & 3.3642 & 0.8329 & 0.2771 \\ 
  Placebo & Other NSAID & 1.5663 & 0.1729 & 0.1212 \\ 
  Placebo & Other NSAID & 23.2551 & 2.3979 & 0.4299 \\ 
  Placebo & Other NSAID & 2.7881 & 1.9459 & 0.9710 \\ 
  Placebo & Other NSAID & 0.0098 & 0.3417 & 0.1733 \\ 
  Placebo & Other NSAID & 0.8146 & 0.4473 & 0.1360 \\ 
  \bottomrule
  \end{tabular}
\end{table}

\begin{table}[ht]
  \centering
  \caption{Study‐level contributions to $\qh$ and estimates of case study 2}
  \label{tab:het_decomp2}
  \begin{tabular}{llccc}
    \toprule
    Treatment 1                      & Treatment 2                        & $\qh^i$     & $y_i$ & $s_i$  \\
    \midrule
Usual Care & Education & 0.1302 & -0.0506 & 0.9300 \\ 
  Usual Care & Education & 0.5265 & 0.5774 & 0.4031 \\ 
  Usual Care & Education & 0.0105 & 0.3082 & 0.2272 \\ 
  Usual Care & Education & 0.4276 & 0.0554 & 0.3510 \\ 
  Usual Care & Education+LCFE & 0.2794 & 0.5096 & 0.2763 \\ 
  Usual Care & Education+LCFE & 2.2587 & 1.8365 & 0.7857 \\ 
  Usual Care & Education+LCFE+HSI & 0.8849 & 0.1665 & 0.2404 \\ 
  Usual Care & Education+LCFE+HSI & 5.9424 & 2.9919 & 1.0663 \\ 
  Usual Care & Education+LCFE+HSI & 3.8729 & 2.7726 & 1.2093 \\ 
  Usual Care & Education+LCFE+Fitting & 0.5618 & 0.1054 & 0.6267 \\ 
  Usual Care & Education+LCFE+Fitting & 0.0364 & 0.6055 & 0.1594 \\ 
  Usual Care & Education+LCFE+Fitting+HSI & 0.7209 & 1.9731 & 0.1851 \\ 
  Usual Care & Education+LCFE+Fitting+HSI & 3.2636 & 1.1045 & 0.3938 \\ 
  Education & Education+LCFE & 0.0000 & 0.8303 & 0.8972 \\ 
  Education & Education+LCFE+Fitting & 0.0000 & 2.2925 & 0.5258 \\ 
  Education+LCFE & Education+LCFE+HSI & 0.0000 & -0.2007 & 0.4945 \\ 
  Education+LCFE+HSI & Education+HSI & 0.0012 & -0.0000 & 0.8165 \\ 
  Education+LCFE+HSI & Education+HSI & 4.5323 & -3.1460 & 1.4647 \\ 
  Education+LCFE+HSI & Education+HSI & 0.0628 & 0.0168 & 0.1775 \\ 
  Education+LCFE+HSI & Education+LCFE+Fitting+HSI & 0.0000 & 1.5730 & 0.0985 \\ 
   \hline    \bottomrule
  \end{tabular}
\end{table}

\begin{table}[ht]
  \centering

  \caption{Study‐level contributions to $\qh$ and estimates of case study 3}
  \label{tab:het_decomp3}
  \begin{tabular}{llccc}
    \toprule
    Treatment 1                  & Treatment 2           & $\qh^i$   & $y_i$ & $s_i$  \\
    \midrule
Placebo/Standard Care & Tocilizumab & 2.0749 & 2.0251 & 0.3834 \\ 
  Placebo/Standard Care & Tocilizumab & 2.6545 & 2.3225 & 0.5215 \\ 
  Placebo/Standard Care & Tocilizumab & 5.4928 & 2.1887 & 0.3055 \\ 
  Placebo/Standard Care & Tocilizumab & 0.5174 & 2.0034 & 0.7376 \\ 
  Placebo/Standard Care & Tocilizumab & 10.7944 & 0.7777 & 0.2116 \\ 
  Placebo/Standard Care & Golimumab & 1.9305 & 1.1393 & 0.4199 \\ 
  Placebo/Standard Care & Golimumab & 2.4465 & 1.3218 & 0.4896 \\ 
  Placebo/Standard Care & Golimumab & 2.5266 & 0.1674 & 0.2444 \\ 
  Placebo/Standard Care & Infliximab & 2.6611 & 3.1922 & 1.4281 \\ 
  Placebo/Standard Care & Infliximab & 1.0493 & 0.6947 & 0.1638 \\ 
  Placebo/Standard Care & Infliximab & 1.8109 & 1.2169 & 0.2633 \\ 
  Placebo/Standard Care & Adalimumab & 7.2287 & 1.7003 & 0.3575 \\ 
  Placebo/Standard Care & Etanercept & 10.2708 & 2.4785 & 0.5199 \\ 
  Placebo/Standard Care & Adalimumab & 3.0538 & 2.0118 & 0.7282 \\ 
  Placebo/Standard Care & Adalimumab & 1.5428 & 1.5067 & 0.6179 \\ 
  Placebo/Standard Care & Adalimumab & 2.9773 & 2.5041 & 1.0229 \\ 
  Placebo/Standard Care & Adalimumab & 5.0636 & 0.3674 & 0.1652 \\ 
  Placebo/Standard Care & Etanercept & 1.2280 & 2.4402 & 1.4690 \\ 
  Placebo/Standard Care & Etanercept & 0.1079 & 0.8722 & 0.1820 \\ 
  Placebo/Standard Care & Etanercept & 2.3753 & 2.4086 & 1.0357 \\ 
  Placebo/Standard Care & Etanercept & 36.3966 & -0.5043 & 0.2182 \\ 
  Placebo/Standard Care & Etanercept & 12.7119 & 1.5416 & 0.2045 \\ 
  Placebo/Standard Care & Certolizumab & 0.2705 & 1.7812 & 0.4261 \\ 
  Placebo/Standard Care & Certolizumab & 0.9151 & 2.9715 & 1.0126 \\ 
  Placebo/Standard Care & Certolizumab & 0.1654 & 2.6022 & 1.4739 \\ 
  Placebo/Standard Care & Abatacept & 3.4401 & 1.5889 & 0.6195 \\ 
  Placebo/Standard Care & Abatacept & 6.3362 & 2.2983 & 0.7383 \\ 
  Placebo/Standard Care & Abatacept & 42.7287 & -1.2171 & 0.2535 \\ 
  Placebo/Standard Care & Abatacept & 12.8475 & 1.1026 & 0.1849 \\ 
  Placebo/Standard Care & Rituximab & 2.3015 & 0.1115 & 0.2875 \\ 
  Placebo/Standard Care & Rituximab & 1.1974 & 1.3838 & 0.7642 \\ 
  Placebo/Standard Care & Rituximab & 3.0328 & 1.3363 & 0.4529 \\ 
    \bottomrule
  \end{tabular}
\end{table}

\begin{table}[ht]
\centering
\caption{Study‐level contributions to $\qh$ and estimates of case study 4}
\begin{tabular}{llrrr}
\toprule
Treatment1  & Treatment2  & $\qh^i$    &  $y_i$    & $s_i$    \\
\midrule
Control & Apixaban & 0.0097 & -0.2335 & 0.0934 \\ 
  Control & Apixaban & 2.5688 & -2.6768 & 1.5187 \\ 
  Control & Dabigatran 150 mg & 0.0487 & -0.8594 & 2.0050 \\ 
  Control & Dabigatran 110 mg & 0.0000 & -0.0910 & 0.1042 \\ 
  Control & Dabigatran 150 mg & 0.0002 & -0.4157 & 0.1133 \\ 
  Control & Rivaroxaban & 0.0000 & -0.1330 & 0.0853 \\ 
\bottomrule
\end{tabular}
\label{tab:het_decomp4}
\end{table}
\clearpage

\setcounter{page}{1}
\setcounter{section}{0}
\setcounter{table}{0}
\setcounter{figure}{0}
\setcounter{equation}{0}

\makeatletter
\renewcommand \thesection{S\@arabic\c@section}
\renewcommand \thetable{S\@arabic\c@table}
\renewcommand \thefigure{S\@arabic\c@figure}
\renewcommand \theequation{S\@arabic\c@equation}
\makeatother

\clearpage

\end{document}